\documentclass{article}

\usepackage{amssymb,amsfonts,amsmath}
\usepackage{cite,enumerate,float}
\usepackage{color}
\usepackage{tikz}
\usetikzlibrary{arrows,snakes,backgrounds}

\def\be{\begin{eqnarray}}
\def\ee{\end{eqnarray}}
\def\nn{\nonumber}

\def\p{\partial}

\def\s{\theta}

\definecolor{red}{rgb}{1,0,0}
\definecolor{orange}{rgb}{1,0.5,0}
\definecolor{violet}{rgb}{0.7,0,1}

\hoffset= - 1.0in         

\begin{document}

\hfill MIPT/TH-08/22

\hfill ITEP/TH-09/22

\hfill IITP/TH-11/22

\bigskip

\centerline{\Large{
Splendeurs et misères of Heavisidisation
}}

\bigskip

\centerline{ {\bf V.Dolotin and A.Morozov}}

\bigskip

\centerline{\it MIPT, ITEP \& IITP, Moscow, Russia}

\bigskip

\centerline{ABSTRACT}

\bigskip

{\footnotesize
Machine Learning (ML)
is applicable to scientific problems,
i.e. to those which have a well defined  answer,
only if this answer can be brought to a peculiar form ${\cal G}: X\longrightarrow Z$
with ${\cal G}(\vec x)$ expressed as a combination of iterated Heaviside functions.
At present it is far from obvious, if and when such representations exist,
what are the obstacles 
and, if they are absent, what are the ways to convert the known formulas  into this form.
This gives rise to a program of reformulation of ordinary science
in such terms -- which sounds like a strong enhancement of the
constructive mathematics approach,
only this time it concerns all natural sciences.
We describe the first steps on this long way.
}

\bigskip

\bigskip

\section{Introduction}

Machine Learning (ML) is a powerful  method to solve classification problems,
i.e. to construct the mappings  $X\longrightarrow Z$ from a set of
sample maps.
This is used to find the answer "by analogy", and has wide applications
to {\it image recognition} and {\it decision taking} problems,
which cover most of needs of the everyday life.

It is natural that people try to extend the range of applications of ML
and include the true scientific problems in it,
what could promote computer applications to science far beyond the
ordinary big data analysis and computer experiments,
often designated as {\it computer physics}.
See \cite{2205.04445} for the latest publication in arXiv and for
the chain of references therein.
However, there is a natural obstacle to this project,
at least in the framework of the {\it steepest descent} method,
included into the current methodology of ML.

In \cite{2204.11613} we formulated an applicability condition for ML
to scientific problems, i.e. to those which have a well defined ("objective") answer:
{\bf this answer should have a representation ${\cal G}: X\longrightarrow Z$
with ${\cal G}(\vec x)$ expressed as a combination of iterated Heaviside functions}.
Then the steepest descent method is applied to the coefficients/parameters $w$,
which can be easily introduced into this formula:
the lifting
\be
{\cal G}(\vec x) \longrightarrow  G(\vec x|w)
\label{lift}
\ee
is straightforward.
It is not uniquely defined and depends on the number of parameters $w$
one wishes to introduce and release.
In general, there is a {\it gauge invariance}, acting on $w$ in $G(\vec x|w)$,
and one needs to fix the gauge in order to eliminate the extra parameters
Also in transition to $G$ one substitutes the sharp Heaviside function
by smoothed $\sigma$-functions, what helps to reduce the gauge freedom and
is more convenient for practical computer calculations.
ML algorithm is to minimize w.r.t. parameters $w$  the functional
\be
{\cal L}(w) := \frac{1}{2}
\sum_\alpha \left| Z^{(\alpha)} - G\Big(\vec X^{(\alpha)}\Big|\,w\Big)\right|^2
\label{func}
\ee
for a given set of association-inspiring sample maps/plots 
$\Big(\vec X^{(\alpha)}, Z^{(\alpha)}\Big)$, enumerated by $\alpha$,
i.e. taking the large-time asymptotics of the solution to
\be
\dot w = -\frac{\p {\cal L}}{\p w}
\label{Lange}
\ee
Our exact ${\cal G}$ is then supposed to arise after the substitution of
the optimal values $\bar w$:
\be
{\cal G}(\vec x) = \lim_{\sigma\rightarrow \theta}G(\vec x|\bar w)
\ee
Usually one does not care about the choice of kinetic term at the l.h.s.
of the Langevin equation (\ref{Lange}) and it breaks gauge invariance.
The stable point and the answer  ${\cal G}(\vec x)$ are gauge invariant in any case. 

Of course, to get a new knowledge, one can not start from a given function ${\cal G}$,
but our goal at this stage is more limited and, in a sense, {\it inverse} --
to understand, what kind of the already {\it known answers} can be reproduced by the
steepest descent ML method.
A necessary step for doing this is expression of these answers
through  iterations of Heaviside functions --
for the purposes of this paper we
use a word "Heavisidisation" to denote this reformulation in terms
of "acceptable" formulas.

In this paper we further elaborate on the elementary examples of \cite{2204.11613}
and attempt to extend them further.
One of the goals (not fully achieved yet) is Heavisidisation of algebraic numbers.
The question is if one can express such numbers in terms of the coefficients
of the underlying polynomial equations
(like Cardano formulas and less explicit constructions for roots and/or discriminants
of more complicated systems of equations)
by using only the iterations of Heaviside functions.

In sec.\ref{elem} we begin the systematic description of elementary operations --
the building blocks for more complicated constructions,
like solution of quadratic equation in sec.\ref{quad}.
After that in sec.\ref{lifting} we consider the lifting (\ref{lift}),
emerging freedom (gauge invariance) and the ways to fix it.
Sec.\ref{prog} contains a brief description of a real program
and its (limited) relation to the true steepest descent method.
It can help to appreciate the amount of human art, which needs to be applied
even in the nearly trivial situations.
The concluding sec.\ref{conc} returns to the philosophic level
and discusses the issues which still remain to be put in a clear form
at the next steps of the Heavisidisation program.

\section{Elementary Heavisidisation
\label{elem}}

\subsection{Heaviside as logical and $\&$ or}

Since Heaviside function $\s(x)$ is going to play the central role in our analysis,
we list here its basic properties and their interpretation in traditional terms
of pure science.

We define Heaviside function
\be
\s = \left\{\begin{array}{ccc} 1 & \text{if} & x>0 \\
0 & \text{if} & x\leq 0 \end{array}\right.
\label{thetadef}
\ee
Then
\be
\s^{\circ 2}(x) :=\s\big(\s(x)\big) = \theta(x)
\ee
Note that we find it convenient to define $\theta(0)=0$.

Further, this function realizes logical operations {\bf AND} and {\bf OR}:
\be
\begin{array}{ccccc}
\wedge(a, b):=\s\big(\s(a) + \s(b) - 1\big) = 1  & \ \text{only if} \
&  \s(a)=1  & {\bf AND} &  \s(b) = 1
\\ \\
\vee(a,b):=\s\big(\s(a) + \s(b)\big) = 1 & <=>
&  \s(a)=1  & {\bf OR} &  \s(b)=1
\end{array}
\label{andor}
\ee

\subsection{Arithmetic operations}

Together with the results of \cite{2204.11613},
by now we have the following set of network building blocks (which may be used/combined as sub-networks):

\begin{itemize}
\item
Logical operations $\wedge,\vee$ on inputs
\item
$\delta_n(x)$: 1 iff $x = n$, defined similar to zero of $(x - n)$, i.e. for integer $x$
\be
\delta_n(x) = \s(x-n+1)-\s(x-n)
\ee
\item
Identity
\be
x=I(x):=\sum_{i=0}^\infty\s(x - i) - \sum_{i=0}^\infty\s(-x - i)
=\sum_{i\in\mathbb{Z}}(\text{sgn}\  {i})\cdot\s\Big((\text{sgn}\ i)(x - i)\Big)
\label{eval_map}
\ee
for integers,
which is easily promoted to rationals
i.e. to an arbitrarily good approximation for reals.
It can be used to construct addition and
multiplication of inputs:\footnote{
Note that we changed the definition of $\s(x)$ from   \cite{2204.11613} to (\ref{thetadef}),
which makes it more functorial  --
thus the small modifications of formulas for (\ref{add})  and (\ref{mult}) 
as compared to that paper. }
\item Addition
\be
x+y
= I(x)+I(y)
\label{add}
\ee
For positive integers addition looks simpler:
\be
x+y = I(x)+I(y)
= \sum_{i=0}^\infty \s(x-i) + \sum_{j=0}^\infty \s(y-j)
\label{addpos}
\ee
\item Multiplication
\be
x\cdot y = \sum_{i,j}^\infty \s\Big(\s(x-i)+\s(y-j) -1\Big)=\sum_{i,j}^\infty \wedge(x-i, y-j)
\label{mult}
\ee
Using these blocks we can also define
\item  Subtraction
\be
x-y = I(x) - I(y)
\label{subtr}
\ee
and approximate
\item
Division
\be
x/y=\sum_{i,j}\wedge\!\left(x-i,\ \frac{\delta_j(y)}{j}\right)
\label{t_div}
\ee
(which is worth comparing to multiplication expression (\ref{mult})).

\item
Square root and roots of other degrees:
\be
x^{1/n} = \sum_{i=0}^\infty \s(x-i^n)
\label{power}
\ee

\end{itemize}

The specifics (or value) of expressions (\ref{add})-(\ref{mult}) is that the numeric operations on the left-hand side are approximated by the network of nodes for which the individual node output {\bf in principal} may not be proportional to the input (the range of $\s$ is $[0,1]$).

One can be surprised: what is the true meaning of (\ref{add})?
The point is that it expresses ordinary addition through operational/network addition.
Note that (\ref{mult}) uses the same operation add, there is no {\it a priori} network multiplication.

\begin{figure}
\begin{center}
\includegraphics[scale=0.4]{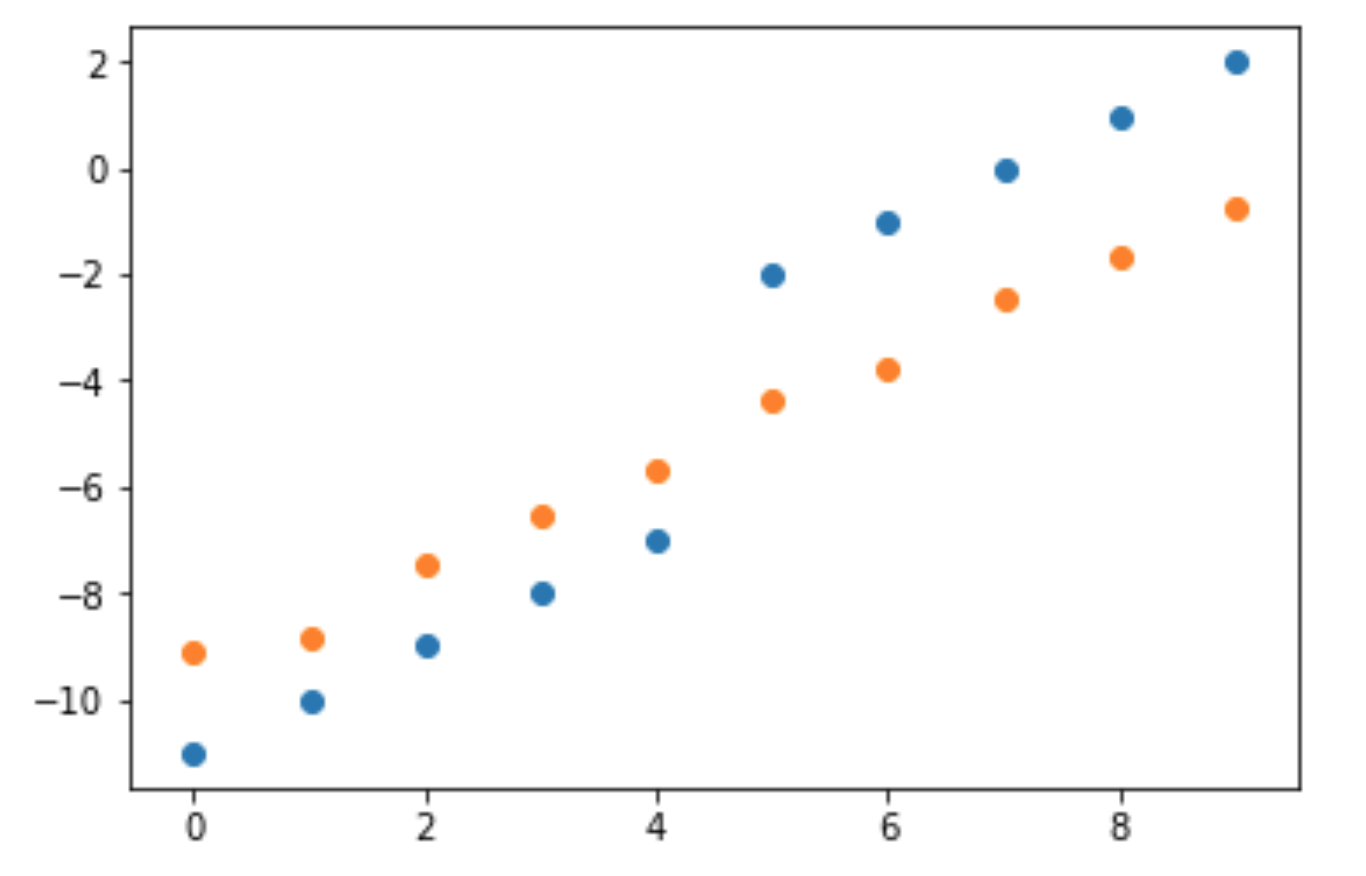}
\caption{The ({\it node number,\,bias value}) plot of descent of initial bias values (blue) to the line (orange) expected in (\ref{eval_map}) via TensorFlow training with ’sigmoid’ activation function (smoothed version of $\theta$).}
\end{center}
\label{fig:bias_eval}
\end{figure}

Figure 1 illustrates the stability of (\ref{eval_map}) as a retraction domain for the steepest descent training procedure. Blue dots show the position of initial bias values, and orange are the values after training, apparently  tending to the line corresponding to bias values in $\sum_{i=0}^\infty\s(x - i)$ of (\ref{eval_map}). The corresponding gradients calculation is discussed in Section \ref{eval_grad}.

\subsection{Zeros detection}

In  \cite{2204.11613} we suggested a formula for a zero of a function $f(x)$: roughly,
\be
\text{zero of}\ f =
 \int_{dx}  x \frac{d}{dx}\theta\Big(f(x)\Big)
\label{zerof}
\ee
It is directly applicable to monotonically growing function with a single zero.
In this section we present more accurate formula without these restrictions
and its generalization to many variables.

Formulas like (\ref{zerof}) can be considered as Heavisidisation of the
equation solving -- but they are not the "clever" formulas for the answers,
like formulas of Cardano type of, say, Koshul-complex constructions of \cite{GKZ,NLA}.
The latter ones require there own Heavisidisation, which we begin to describe
in sec.\ref{quad} below.

\subsubsection{1D}

For a lattice approximation $(f_i)$ of a continuous function $f$ denote
\be
\delta_i(f):=
\vee(\s(f_{i+1})-\s(f_{i}),\ \s(f_{i})-\s(f_{i+1}))
=\s\left(\s\big(\s(f_{i+1})-\s(f_{i}))+\s(\s(f_{i})-\s(f_{i+1})\big)\right)
\label{deltaf}
\ee
the indicator function being 1 iff $f$ has (odd number of) $0$ somewhere between the nodes $i$ and $i+1$.
The union operation is used to take care of the gradient direction and handle both decreasing
and increasing behavior in the vicinity of zero.
Degenerate zeroes are not accurately described in this lattice approximation.

\subsubsection{2D}

The indicator function, which checks if a function $f$ on 2D lattice
has 0 within the square $(i,j),(i+1, j+1)$ is
\begin{align}
\delta_{i,j}(f)=&\vee(\delta_i(f_{\bullet j}),\delta_j(f_{i\bullet}))=\s(\s(\delta_i(f_{\bullet j}))+\s(\delta_j(f_{i\bullet})))
\cr\overset{(\ref{deltaf})}{=}&
\s\left(\s\left(
\s\left(\s\big(\s(f_{i+1,j})-\s(f_{i,j}))+\s(\s(f_{i,j})-\s(f_{i+1,j})\big)\right)
\right)\right.
\cr+&\left.\s\left(
\s\left(\s\big(\s(f_{i,j+1})-\s(f_{i,j}))+\s(\s(f_{i,j})-\s(f_{i,j+1})\big)\right)
\right)\right)
\end{align}
where $f_{\bullet j}$ denotes the restriction of the two-variables function onto 1-dimensional sub-lattice with fixed $j$ (similar to $f(\cdot, y)$). The union is used to take care of the case when $f$ is constant along one of the lattice axes, $i$ or $j$.

The indicator function checking if the functions $f,g$ on 2D lattice both have 0 within the square $(i,j),(i+1, j+1)$ is
\be
\delta_{i,j}(f,g)=\wedge(\delta_{ij}(f),\delta_{ij}(g))=\s(\s(\delta_{ij}(f))+\s(\delta_{ij}(g))-1)
\ee

The position of a joint zero of $f,g$ may be approximated by the following network with 2-dimensional output

\be
\left(\sum_{ij}\frac{i}{N}\delta_{i,j}(f,g),\ \ \sum_{ij}\frac{j}{N}\delta_{i,j}(f,g)\right)
\ee

\subsubsection{Examples}

Relations like (\ref{zerof}) are trivial if we use $\s'(x) = \delta(x)$
but get a little tricky in other terms, especially in the difference form.
In linear case $F(x)=x-b$ we can apply integration by parts:
\be
b = \int_{-\Lambda}^\Lambda x\s'(x-b)dx = \left. x\s(x-b)\right|_{x=-\Lambda}^\Lambda
- \int_{-\Lambda}^\Lambda s(x-b) dx = (\Lambda - 0) -(\Lambda-b) = b
\ee
for sufficiently big ultraviolet cutoff $\Lambda > |b|$.
In terms of discrete summation
\be
b = \sum_{i=-\Lambda}^\Lambda \frac{i}{N}
\left\{\s\left(\frac{i+1}{N}-b\right) - \s\left(\frac{i}{N}-b\right)\right\}
= \frac{\Lambda}{N}\cdot \s\left(\frac{\Lambda+1}{N}-b\right)
  +  \sum_{i=-\Lambda}^\Lambda \left(\frac{i-1}{N}-\frac{i}{N}\right)\s\left(\frac{i}{N}-b\right)=
\nn \\
= \frac{\Lambda}{N}\cdot \s\left(\frac{\Lambda+1}{N}-b\right)
  - \frac{1}{N}  \sum_{i=-\Lambda}^\Lambda \s\left(\frac{i}{N}-b\right)
=\frac{\Lambda }{N} -\left(\frac{\Lambda}{N}-b\right)    = b  \ \ \ \ \
\ee
(in fact there is a $1/N$ correction for this choice of discretisation).

\bigskip

Likewise
for quadratic case we get
\be
\sum_{i=-\infty}^\infty \frac{i}{N}\cdot \left\{\s\left(
\frac{i+1}{N}b +c+\left(\frac{i+1}{N}\right)^2\right)
- \s\left(\frac{i}{N}b +c + \left(\frac{i}{N}\right)^2\right)\right\}
\ee
This corresponds to the network with 2 inputs, $b$ and $c$.
We can make further transforms to reproduce (\ref{power}) with $n=2$:
{\footnotesize
\be
=\lim_{\Lambda\to\infty}\sum_{i=-\Lambda}^\Lambda \frac{i}{N}\cdot \left\{\s\left(
c-\left(\frac{b}{2}\right)^2+\left(\frac{i+1}{N}+\frac{b}{2}\right)^2
\right)
- \s\left(
c-\left(\frac{b}{2}\right)^2+\left(\frac{i}{N}+\frac{b}{2}\right)^2
\right)\right\}\overset{D:=b^2-4c}{=}\cr
=
\frac{\Lambda}{N}\left.\s\right|_{i=\Lambda}-\frac{-\Lambda}{N}\left.\s\right|_{i=-\Lambda}+\sum_{i=-\Lambda}^\Lambda \frac{i-1}{N}\cdot \s\left(
-D/4+\left(\frac{i}{N}+\frac{b}{2}\right)^2
\right)
- \sum_{i=-\Lambda}^\Lambda\frac{i}{N}\cdot\s\left(
-D/4+\left(\frac{i}{N}+\frac{b}{2}\right)^2
\right)\cr
=
\frac{2\Lambda}{N}\left.\s\right|_{i=\Lambda}+\sum_{i=-\Lambda}^\Lambda-\frac{1}{N}\s\left(
-D/4+\left(\frac{i}{N}+\frac{b}{2}\right)^2
\right)
=\frac{2\Lambda}{N}\left.\s\right|_{i=\Lambda}+\sum_{i=-\Lambda}^\Lambda-\frac{1}{N}\s\left(
-D/4+\left(\frac{i+Nb/2}{N}\right)^2
\right)\cr
=
\frac{2\Lambda}{N}\left.\s\right|_{i=\Lambda}
-\frac{1}{N}
\sum_{i=-\Lambda}^\Lambda \s\left(
-D/4+\left(\frac{i'}{N}\right)^2
\right)
\overset{r:=N\sqrt{D/4}}{=}\frac{2\Lambda}{N}\left.\s\right|_{i=\Lambda}-
\frac{1}{N}
\left(\sum_{i=-\Lambda}^{-r}+\sum_{i=-r}^{r}+\sum_{i=r}^{\Lambda}\right)
\s\left(
-D/4+\left(\frac{i'}{N}\right)^2
\right)
\cr
=
\frac{2\Lambda}{N}\left.\s\right|_{i=\Lambda}-
\frac{1}{N}\left(\sum_{i=-\Lambda}^{0}+\sum_{i=1}^{\Lambda}\right)\s(1)+\frac{1}{N}\sum_{i=-r}^r\s\left(
D/4 - \left(\frac{i'}{N}\right)^2
\right)=\frac{1}{N}\sum_{i=-r}^r\s\left(
D/4 - \left(\frac{i'}{N}\right)^2
\right)
\nn
\ee
}
This is a finite sum, which is indeed (according to (\ref{power})) equal to
\be
= 2\sqrt{D/4}=\sqrt{b^2-4c}
\label{diff2desc}
\ee

\noindent
Here at the intermediate steps we introduced notations $D:=b^2-4c,\ r:=N\sqrt{D/4}$. For the steps 2 and 5 of our transformations we use the  translation of $i$.

\subsection{Sector functions}

\subsubsection{Example: 1D}
Take an example when $X=\mathbb{R}^1$ and $Z=\{0, 1\}$, i.e. we classify points of real line as belonging to 2 possible classes. The mapping which we want to approximate by our network looks as:
\be
g:X\to\left\{
\begin{array}{l}
0,\ x < 2 \\
1,\ 2\le x\le 3 \\
0,\ x>3
\end{array}\right.
\ee
i.e. points in the segment $[2,3]$ belong to the class $1$, and the rest to $0$.

Lets take a level-1 network. The output value has the general form
\be
G=\sum_iw_1^i\sigma(w_0^ix + \xi_0^i)+\xi_1
\ee
If we take $\sigma$ to be a Heaviside function,
then we can indeed express the well known answer in the following form:
\be
{\cal G}(x)=\theta(x-2) - \theta(x-3) + 0
\label{class_1d}
\ee
This means that we can make tie network of layer-1 with 2 cells
which is going to converge to the values:
\be
\bar w_0^1=\bar w_0^2=1,\ \bar \xi_0^1 =-2,\ \bar \xi_0^2 =-3
\nn\\
\bar w_1^1=-\bar w_1^1=1
\ee

{\it One can see that this way we can  actually express
any step function on the line, which solves the classification problem
for the 1-dimensional configuration space $X$.}

\subsubsection{Example: 2D}

For simplicity consider not an arbitrary and limited domain,
but just a sector on the plane $X=\mathbb{R}^2$.
This means that the mapping we want to
describe is the characteristic function of a sector:
\be
g:(x^1,x^2)\mapsto\left\{
\begin{array}{lcc}
1 &{\rm for} \  x^1, x^2>0\\
0 &   {\rm otherwise}
\end{array}\right.
\ee

Solution is provided by the level-2 network.
In general for such network
\be
G = \sum_jw_2^j\s(\sum_iw_1^{ji}\sigma(w_0^{i1}x_1+w_0^{i2}x^2+\xi_0^i)+\xi_1^j)+\xi_2
\ee

The approximation  to our characteristic function gets exact for $\sigma=\theta$:
\be
{\cal G}(x_1,x_2) =-\theta\Big(\theta(-x_1)+\theta(-x_2)-1\Big)+1
\ee
When both $x_1$ and $x_2$ are positive, the argument of external $\theta$ is $-1$, and $y=1$.
When any of $x_1$ or $x_2$ is negative the argument is $0$ or $1$, and $y=0$.

Thus the relevant network is layer-2 with 2+1 cells and converges to
\be
\bar w_0^{11}=\bar w_0^{22}=-1 \nn \\
\bar w_1^{11}=\bar w_1^{12}=1 \nn \\
\bar w_2^1=-1,\ \bar \xi_2=1
\ee

{\it Combination of sector characteristic functions may provide an approximation
(with precision depending on the number of network cells) to any step function on the plane.}

\subsubsection{General classification/approximation problem}

Let us take a "modified" Heaviside function with $\s(0)=0$.
Then it becomes a well-defined operation with the property
\be
\s\circ\s=\s
\label{u_p}
\ee

For approximation of the characteristic of $(n+1)$-dimensional sector
$$x_0,x_1,\dots,x_n>0$$
we have the 2-layer network
\be
{\cal G}(x)=\s\left(\sum_{i=0}^n\s(x_i) - n\right)
\label{class_g}
\ee

Note, that the property (\ref{u_p}) allows us to re-write 1D example above as a particular case of 2-layer formula (\ref{class_g}). For 1-d sector $x_0>0$, i.e. for $n=0$:
$${\cal G}(x)=\s(\s(x_0))$$
and then characteristic (\ref{class_1d}) of the segment $[2,3]$ looks as:
$${\cal G}(x)=\s\big(\s(x-2)-\s(x-3)\big)$$
{\it Since a general multivariate step function may be approximated by (a linear combination of) characteristics of sectors and a any multivariate function may be approximated by (a linear combination of) step functions then 2-layer networks provide a solution for a general approximation problem.}

\subsection{Non-Picassian networks}

Using sectoral functions as building blocks for approximation makes it pretty adequate for approximating step functions whose plots would look as a sample of spacial cubism art. This apparent limitation is due to the fact that in the Heaviside operators of layer $n+1$ we use linear combination of values of layer $n$. Having that formulated immediately suggests some generalizations.

\subsubsection{The role of bias}

Lets add to each layer $l$ of our network an extra node $\nu_l$. And lets make out of those nodes a trivial "identity" sub-network
$$\nu_0\equiv x_0=1,\ \ \ \ \  \nu_l = \s^{\circ l}(\nu_0)=1$$

Then our initial general form network with $n$-dimensional input $(x_1,\dots,x_n)$
\be
{\cal G}_{j_{l+1}}(x)=
\sum_{j_l=1}^{n_l}W_{l}^{j_{l+1}j_{l}}\cdot\s\!\left(\dots \
\s\!\left(\sum_{j_1=1}^{n_1}W_1^{j_2j_1}\cdot
\s\Big(\sum_{i=1}^nW_0^{j_1i}x_{i}+\xi_0^{j_1}\Big)
+\xi_1^{j_2}\right)\dots+\xi_l^{j_{l+1}}\right)
\label{n_wb}
\ee
may be rewritten using the extension by the identity sub-network in homogeneous form as
\be
\sum_{j_l=0}^{n_l}W_{l}^{j_{l+1}j_{l}}\cdot\s\!\left(\dots\s\!\left(\sum_{j_1=0}^{n_1}W_1^{j_2j_1}\cdot
\s\Big(\sum_{i=0}^nW_0^{j_1i}x_{i}\Big)\right)\dots\right)
\label{n_nb}
\ee

where all the bias terms are replaced by links to the identity sub-network
$$W_l^{j0}\nu_l = b_l^{j}$$

After adding that extension to the layer, we may certainly omit the restrictions for the values of $\nu_l$ and $W_l^{j0}$ when applying the ML process on  computer, just remembering that our network with bias (\ref{n_wb}) is a particular case of (\ref{n_nb}).

\subsubsection{World getting smoother}

The next generalization step would be to use a higher degree expressions of output $Y_{(l)}$ from layer $l$ for the input $X_{(l+1)}$ for the layer $l+1$. Using more common tensor algebra notations and assuming summation for repeated indexes:

\be
x^j=W_{i_1\dots i_d}^{(l),j}y^{i_1}\dots y^{i_d}
\label{n_np}
\ee

Denote $x^I:=x^{i_1}\cdot\ldots\cdot x^{i_d}$. Then using multi-index notations we can write our network in a rather symbolic (but still recoverable to the original notations) form:
\be
{\cal G}_{j_{l+1}}(x)=
W_{J_{l}}^{(l),j_{l+1}}\cdot\s\!\left(\dots W_{J_2}^{(2),J_3}\cdot\s\!\left(W_{J_1}^{(1),J_2}
\cdot \s\Big(W_{I}^{(0),J_1}x^I)\Big)\right)\dots\right)
\ee

As in the case of bias, we can note that the network (\ref{n_nb}) is a specialization of (\ref{n_np}) to the particular case of
\be
y^0=\nu_l=1
\ee
and
$$W_{i_1\dots i_d}^{(l),j} = 0$$
for more then 1 non-zero indexes in $i_1\dots i_d$. To go to the generalization it is enough just to omit this restriction in concrete calculations.

The use of the above form of connection between layers dramatically changes the opportunities of approximating step functions with smoother support boundaries. For instance to get an exact presentation of a characteristic function of a 2D ellipse
$$ax_1^2+bx_2^2 < 1$$
it is enough to use a 1-layer network with a single node
\be
\s(\nu_0^2-(ax_1^2+bx_2^2))
\ee
(remember $\nu_0\equiv 1$)while the approximation by sectoral functions would need a 2-layer network with tens or hundreds of nodes (depending on the required precision).

\subsubsection{Characteristic functions of bounded  domains}

In fuller generality,
a domain $F(\vec x)< 1$ has an {\it acceptable} characteristic function
\be
\s\Big(1-F(\vec x)\Big)
\ee
If we have a domain, which is unit of two, $F(\vec x)<1$ and $G(\vec x)<1$,
then we can use (\ref{andor})
to claim that the characteristic function is given by a two-layer expression
\be
\s\left(\s\Big(F(\vec x) + G(\vec x)\Big)\right)
\ee
if it is their intersection then the characteristic function is
\be
\s\left(\s\Big(F(\vec x) + G(\vec x)-1\Big)\right)
\ee
Can we also describe the boundary?

\section{Towards algebraic numbers
\label{quad}}

\subsection{Solving quadratic equation\label{quad_eq}}

Now we can attempt to solve the first {\it non-trivial} problem:
quadratic equation. The alternated sum of zeros $\int\frac{\partial\theta(F(x))}{\partial z}zdz$ in case of $F$ being quadratic polynomial has a concise analytic form (as a specific case of Vandermonde determinant)
\be
\int \frac{\p \s(z^2+bz+c)}{\p z}\,z dz = \sqrt{b^2-4c}
\label{quadeq}
\ee
Hevisidization of both parts looks as
\begin{align}
\sum_{i=-\infty}^\infty \frac{i}{N}\cdot &\left\{\s\left(\left(\frac{i+1}{N}\right)^2
+  \frac{i+1}{N}\,b +c\right)
- \s\left(\left(\frac{i}{N}\right)^2 + \frac{i}{N}\,b +c\right)\right\} = \int \frac{\p \s(z^2+bz+c)}{\p z}\,z dz
\cr=& \sqrt{b^2-4c} \overset{(\ref{power})}{=} \frac{1}{N} \sum_{i-0}^\infty \s\left(
b^2-4c
- \left(\frac{i}{N}\right)^2\right)
\cr\overset{(\ref{mult})}{=}&
\frac{1}{N} \sum_{i-0}^\infty \s\left(  \sum_{j_,k}\s\Big(\s(b-j)+\s(b-k)-1\Big)-4c
- \left(\frac{i}{N}\right)^2\right)
\label{root2eq}
\end{align}
At the r.h.s. we consider $-4c$ just as a number, to simplify the formula --
it is easy the include it by the rules of sec.\ref{elem}, but this will raise
the number of layers, what will distract us from the main line of reasoning at this moment.

When learning, the net can discover any side of the equation (\ref{root2eq}).
However, one (left) is trivial definition of the difference of zeroes,
while the other (right) is an important scientific formula.
How to teach the net to distinguish?
And to discover the r.h.s.?

Note, that without knowing the originating equation (\ref{quadeq}) we could prove the equality of both ends of (\ref{root2eq}) following (\ref{diff2desc}).

\subsection{Ambiguity of Heavisidisation
\label{ambig}}

In fact, already $b^2$ can be defined in different ways, e.g.
\be
b^2 = \sum_{i,j=0}^\infty\s\Big( \s(b-i) + \s(b-j)-1\Big)
\ee
or
\be
b^2 = \sum_{i=0}^\infty \s(b-\sqrt{i})
\label{sqrt}
\ee
In this case it is more difficult to say what is trivial, and what is the law of nature.

\subsection{Architecture of the network}

In computer science the difference between these formulas is associated with the
"architecture" of the network.
Thus we see, that the "laws of nature" instead of "trivialities" arise only for
appropriate architecture choice(?!)
In other words, {\bf the difference has no sense for the computer}(?!) --
at least, it remains to understand, what it is.

Another obvious question is what if we use architecture, sophisticated enough to incorporate both formulas?

\subsection{Cardano formulas for cubic and quartic equations}

The somewhat mysterious use of (\ref{zerof}) in arriving to discriminant
by  means of Heaviside functions in Section \ref{quad_eq} is essentially
due to the fact that for quadratic polynomial the difference of roots
(calculated by (\ref{zerof})) is a particular case of Vandermonde determinant (discriminant).
It is still a question whether we should expect any algebraic outcome in case of degrees $3,4$.

\subsection{Beyond Cardano}

Even more interesting is
what happens if we take polynomial of degree 5 and higher?
The analogue of the l.h.s. of (\ref{root2eq}) continues to exist,
what if there is also a clever r.h.s.
Perhaps, it exists(?),
just "inverse Hevisidization" does not work in a simple way --
the emerging combination of  Heavisides (if any)
can not be written in terms of roots?

\section{Steepest descent method for elementary examples
\label{lifting}}

\subsection{Linear function}

According to (\ref{addpos}) and (\ref{power}),
for
\be
{\cal L}  = \sum_K {\cal L}_K^2 = \sum_K \left(b_K^{1/n} - \sum_i  W_i\cdot\s(b_K-\xi_i^n)     \right)^2
\ee
we need to get the answer (stable point)
\be
\bar W_i=1, \ \ \ \   \bar \xi_i = i^n
\label{addsoln}
\ee
These should come as solutions (limiting points) of  the steepest descent equations
\be
\dot W_i =  -\sum_K {\cal L}_K\cdot \s(b_K-\xi_i^n)
\nn \\
\dot\xi_i = -\sum_K {\cal L}_K\cdot  W_i\,\delta(b_K-\xi_i^n)
\ee
Of course, ${\cal L}_K=0$, which is an identity for (\ref{addsoln}),
provides a solution --
but we need to solve these equations without knowing the answer.
Already for $n=1$ it is a non-trivial task.

Let us specify the problem: restrict it to $n=1$ and to integer $b$.
Then the maximally big teaching set consists of pairs ($K,b_K=K)$.
In this case further reduction trivializes the problem:
if we fix all $\xi_i=i$ then $W_i$ will be defined by minimization of
\be
{\cal L} = \frac{1}{2} \sum_K \left(K - \sum_{i=0}^{K-1} W_i\right)^2
\ee
with a sum over positive integers $K$ (the sample set at $n=1$, $\left\{b_K=K\right\}$).
This obviously  implies $\bar W_i=1$
-- this is what makes all the items  in the sum of squares vanishing.
Still the steepest descent equations (\ref{Lange}) even in this oversimplified case are
rather complicated.
Simple are combinations like
\be
\dot W_0 -\dot W_1 = 1-W_0 \nn \\
\dot W_1 - \dot W_2  = 2-W_0-W_1 \nn \\
\dot W_2 - \dot W_3 = 3-W_0-W_1-W_2 \nn \\
\ldots
\label{sdeqsforn=1}
\ee
(what already calls for a more clever definition of the kinetic term at the l.h.s.)

At the same time, releasing $\xi_i$ is not a truly good idea:
clearly sufficiently small deviations $\xi_i$ from $i$ are allowed,
it is enough that, say, $i-1<\xi_i \leq i$ for all $i$.
One can fix this ambiguity either by going to dense sets of $K$
(like rationals, with appropriate rescaling of  $i$),
or by smoothing the  sharp Heaviside functions $\s(x) \longrightarrow \sigma(x)$.
Thus {\bf one should be careful in precise formulation of the optimisation problem}
and defining the allowed type/range of adjusting parameters.???

Another point is that making the teaching set incomplete,
breaks the validity of the answer irreversibly.
For example, if we omit just the single sample value at $K=2$,
this will eliminate $W_1$ and the second line from (\ref{sdeqsforn=1}).
This, in turn, will shift the solution to $\bar W_3=2$, without changing all the other
$\bar W_0 = \bar W_2 = \bar W_3 = \ldots = 1$.
In other words, in this way we teach that
\be
K \approx \s(K) + 2\s(K-2) + \sum_{i=3}^\infty \s(K-i)
\ee
what is not true for $K=2$,
in variance of correct answer
\be
K=\sum_{i=0}^\infty \s(K-i)
\ee
In other words, the {\bf net does not want to guess the omitted data,
or, better to say, guesses it wrongly}.
One can of course search for a better net, withe better associations,
but this where science (exact knowledge) stops and gives room to an art.

\subsection{A more formal presentation}
\label{eval_grad}

Let us train our 1-level network
\be
y = \sum_i\s(w_ix+\xi_i)
\ee
on 2 input-output $(x, y)$ samples - $(a, a)$ and $(b,b)$, for $a,b$ being positive integers.

The functional to minimize is:
\be
L = \left(\sum_i\s(aw_i+\xi_i) - a\right)^2 + \left(\sum_i\s(bw_i+\xi_i) - b\right)^2=(y(a)-a)^2+(y(b)-b)^2
\ee
Then
\be
\frac{\partial L}{\partial  b_i}=2\delta(\xi_i+aw_i)(y(a) - a)+2\delta(\xi_i+bw_i)(y(b) - b)
\ee

Take the starting network state:
\be
w^0_i = 1,\ \xi^0_i = 0
\ee
Then
\be
\left.\frac{\partial L}{\partial  \xi_i}\right|_{w_i=1,\xi_i=0}
=2\delta(a)\left(\sum_i\s(a) - a\right)+2\delta(b)\left(\sum_i\s(b) - b\right)
\ee
since for $x>0$ we have $\sum_i\s(x)=+\infty>0$ then (assuming for the smoothed variant of $\delta()$ to be everywhere positive)
\be
\left.\frac{\partial L}{\partial  b_i}\right|_{w_i=1,\xi_i=0}>0
\ee
i.e. $\xi_i$ should be diminishing.

Take the starting state:
$$w^0_i=1,\ xi^0_i=-2i$$

Then
$$\sum_i\s(a-2i) - a = \frac{a}{2}-a=-\frac{a}{2}$$
and
\be
\left.\frac{\partial L}{\partial  b_i}\right|_{w_i=1,\xi_i=-2i}
=2\delta(-2i+a)\left(-\frac{a}{2}\right)+2\delta(-2i+b)\left(-\frac{b}{2}\right)
=-(a\delta(-2i+a)+b\delta(-2i+b))<0
\ee
i.e. $b_i$ is supposed to be increasing.

Then it is not surprising that in general our network is going to converge to the state
$$\bar w_i=1,\ \bar \xi_i=-i$$
which corresponds to the "identity" transform $I(x)$.

\subsection{Powers}

We can now switch from linear function to the simplest power.
As a lifting of (\ref{sqrt}) we can take, say,
\be
k^2 = \sum_{i=0}^\infty W_i\s(k-\xi_i)
\ee
The desired stable point (\ref{sqrt}) is
\be
\bar W_i = 1, \ \ \ \ \
\bar\xi_i = \sqrt{i}
\ee
However, there are many more.
Even if one fixes $\xi_i = \sqrt{i}$, what we need to minimize is
\be
{\cal L} = \frac{1}{2} \Big((1^2-W_0)^2 + (2^2-W_0-W_1-W_2-W_3)^2 + \ldots \Big)
\ee
and the fixed points are just
\be
\bar W_0=1
\nn \\
\bar W_0+ \bar W_1 + \bar W_2 + \bar W_3 = 4
\nn \\
\ldots
\ee
what does not allow to fix all the individual $\bar W_i$:
we get $ \bar W_1 + \bar W_2 + \bar W_3 = 3$, not
necessarily   $ \bar W_1 = \bar W_2 = \bar W_3 = 1$.
Releasing $\xi_i$ only worsens the situation.

Note that this is {\it not} the ambiguity, discussed in (\ref{ambig}) --
between the different Hevisidisations of the {\it answer}.
Now we face the problem for a {\it given answer} (fixed point) --
and this is additional ambiguity of the lifting \cite{lift}.

\subsection{Continuous case and the gauge freedom}

The simplest way to analyze this problem is to substitute our
discrete examples by a continuous one with $i$ substitute by a
continuous parameter $z$, say,
\be
g(k) = \int_0^\infty  W(z)\s\Big(k-\xi(z)\Big) dz
\label{contexa}
\ee
Integral is invariant under change of variables.
We can fix this freedom by choosing either $\xi(z)$ of $W(z)$/
or by imposing any desired relation between them.
In other words, there is always a 1-parametric set
of equivalent formulas of the type (\ref{contexa}),
and no one of them is truly distinguished.
This makes the status of particular choice like (\ref{power}),
\be
k^\alpha =  \int_0^\infty   \s\Big(k-z^{1/\alpha}\Big) dz 
\ee
(proved by taking the $k$-derivative) somewhat unclear/shaky.

In other words, there is a gauge invariance, which need to be fixed.
The question is where is this step in the standard formalism,
and what are canonical??? ways of this gauge fixing.
Langevin equation (\ref{Lange}) breaks this gauge invariance,
but it is restored at the fixed points $\bar w$ (extrema of ${\cal L}$).

Additional problem is that the kernel of Heaviside transform
has extra freedom,
\be
\theta(e^{u(z)}z) = \theta(z)
\ee
(just adding no extra zeros, besides $z=0$)
unlike the kernels of, say, Fourier or Mellin transforms.
This additional gauge freedom is  fixed by smoothing
$\theta \longrightarrow \sigma$, ???
but in the study of Heavisidisation it should be also taken into account.  

Iterated Heaviside transforms are less familiar:
\be
g(k) =
\int W_2(z_2)\cdot \s\left\{\int W_1(z_1,z_2)\cdot\s\Big(k-\xi_{12}(z_1,z_2)\Big)dz_1 \right\} dz_2
\ee
and what is the way to find and fix the gauge freedom here?

One can also lift ordinary integrals to matrices (Heaviside matrix models).

Still another issue is that kinetic term at the l.h.s. of (\ref{Lange})
does not respect gauge invariance.
In fact, already the simplest example (\ref{sdeqsforn=1}) demonstrates that (\ref{Lange})
can be not the most adequate choice even without this difficulty.
A clever choice of kinetic term is not needed for practical applications of ML,
but it is clearly desired for a truly scientific approach.

\section{TensorFlow as a commonwealth of artificial and human
\label{prog}
}

Romantically, one imagines artificial intelligence as a program,
which gets a data set as input and produces an answer in the output.
In practice it is often far from this ideal --
especially in application to "scientific problems",
which we discuss in this paper.
Formally, the difficulty is to get onto the "right" orbit,
which converges to the true answer --
if at all exists in the suggested architecture.
So far we concentrated on the last issue --
emphasized the need for the true answer to allow Heavisidisation.
However, in the previous section we saw, that this is not enough --
even if the true answer can be expressed through iteration of
Heaviside functions, the steepest descent method does not
obligatory allow to find it, starting from generic initial conditions.
In this section we briefly remind how the actually-existing programs work
and explain why they are in fact infinitely far from the
artificial-intelligence dream.
The crucial watershed between human and artificial intelligence
is not crossed yet and the crucial decisions are still made by humans.
Computer still only helps.

We take as an example the widespread TensorFlow program and illustrate
 the way it solves our elementary problems, Heavisidised in the previous sections.
 Unlike mathematical formulation (2) assuming existence of some minimizing subspace of network states,
  practical calculation at each step uses for computing the gradient only one of training examples,
  makes a single timid step in the corresponding direction and then iterates
with the rest of samples set.
In other words, instead of making the step satisfying each of examples
it makes a sequence targeting samples successively.


Such procedure allows one not to be occupied with the question of compatibility of gradient descent conditions,
in particular with the fundamental question which size of the network is suitable for having a reasonable solution.
 And as a payback we have no definite idea on when the training procedure should/may be finished
 (delegating it to the 'epochs' parameter),
 or even regarding its potential convergence, which makes the role of human experience
 and a proper training scenario indispensable.

\subsection{Example}

For illustration we take a TensorFlow model (Python code is attached at the end of this section)
which we already used in getting the Fig.1.
However, this time we describe the actual procedure in a more realistic way,
and explain that the nice picture in Fig.1 is in fact a result of a clever {\it human} work,
not a blind application of a well-defined algorithm.
In our oversimplified example "clever" is quite straightforward --
the secret is to use a homogeneous distribution of training examples.
Here we just show what happens, if instead some special/deformed distributions are taken.
Still this illustrates the problem, which is quite general and requires serious human
insight in more sophisticated situations where the criteria of homogeneity are not that obvious.
Of course, this is not a surprise for people familiar with the true applications of ML,
but it is not sufficiently well known and appreciated by the community of pure scientists,
who did not (yet) use ML in their work.

So, we look at the simplest possible 1-layer network
$$\sum_{i=1}^{10}\s(x + b_i)$$
with 10 nodes in the layer.
Making use of above mentioned "gauge invariance" we normalize the initial values of $W$ to 1
and take the initial state of bias as
$$
b_i=\left\{\begin{array}{ll}
i+1, & 1\le i\le 5 \\
i-1, & 6\le i\le 10
\end{array}\right.
$$
being a deformatiion of (\ref{eval_map}). We want to train our network on examples of identity  map $x\mapsto y$, using the set of 2 examples:
$$(x, y) = (2, 2)\ \text{and}\ (5,5)$$

For the first training session we submit those examples to our network in a sequence with inputs $[2,5,5,5,5]$, and for the second session we use the sequence $[2,2,2,2,5]$. In  each session we iterate the submission 2000  (the value of 'epochs' parameter) times.
The resulting values of $b_i$ are shown in Figure 2 as orange dots compared to initial blue.

In case if TensorFlow was doing the descent following the system of equations corresponding to the set of examples (as mathematical formulation suggests) we would certainly arrive in the deterministic way to the same state of $b_i$ in both cases (making the same number of steps from the same initial state). But since each elementary  step is actually made based on  the currently picked example from the sequence provided by a human, it makes our network to travel in the space of states according to our training scenario.

\bigskip

To summarize, we can split the work of TensorFlow into the following steps:

\begin{enumerate}
\item
choose a network configuration, the set of layers
\item
take initial state of $w$ and $b$ (from standard variants or manually)
\item
pick an example according to the training set sequence (provided by human)
\item
compute the descent gradient (for selected example) according to (\ref{func})-(\ref{Lange}) (handled by TensorFlow with possible custom activation function)
\item
change $w$ and $b$ accordingly
\item
iterate steps 3.-5.
\end{enumerate}

Human interference is needed at the steps 1-3.
Moreover, it may be unavoidable -- if machine makes the choice in 2,
or we provide an inappropriate sequence of training example in 3,
the process may never converge to the desired values,
training will not work.

\begin{figure}
\begin{center}
\includegraphics[scale=0.5]{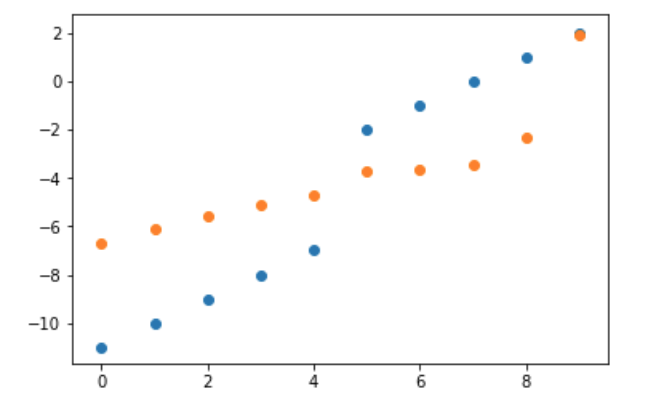}
\includegraphics[scale=0.5]{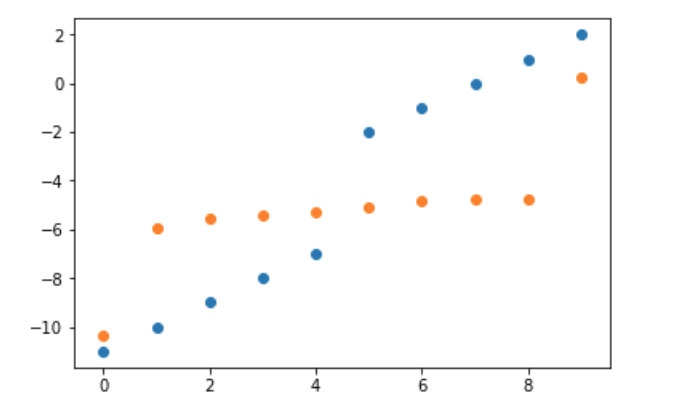}
\caption{The ({\it node number,\,bias value}) plot as a result of TensoFlow model.fit() calculation (orange) with the same initial state (blue) of $\{b_i,\ i=1\dots 10\}$ using training sequences [2, 5, 5, 5, 5] and [2, 2, 2, 2, 5] correspondingly.}
\end{center}
\label{fig:bias_evalseq}
\end{figure}

\subsection{Python listing}

Below is the TensorFlow calculation from example of Section \ref{prog}.

\begin{verbatim}
import tensorflow as tf
from tensorflow.keras import Sequential
from tensorflow.keras.layers import Dense

import math
import random
from matplotlib import pyplot

# NETWORK CONFIGURRATION/INITIALIZATIION
NCell = 10

binit = [None]*NCell
for i in range(NCell):
    mv = 2 if i < NSample/2 else -2
    binit[i] = -i + mv
binit = tf.keras.initializers.Constant(binit)

model = Sequential()
model.add(Dense(NCell, activation = 'sigmoid', kernel_initializer = 'ones', bias_initializer = binit))
model.add(Dense(1, kernel_initializer = 'ones', bias_initializer = 'zeros', trainable=False))
model.compile(loss='mean_absolute_percentage_error', optimizer='adam', metrics=["mse", "mae"])

# TRAINING
NStep = 2000
# training scenario 1
model.fit([2,5,5,5,5,5], [2,5,5,5,5,5], epochs=NStep, verbose=0)
# training scenario 2
#model.fit([2,2,2,2,2,5], [2,2,2,2,2,5], epochs=NStep, verbose=0)

# OUTPUT
abias = sbinit
pyplot.scatter(x, abias)

aweight=model.layers[0].weights[0].numpy()[0]
abias=model.layers[0].weights[1].numpy()
for i in range(NSample):
    abias[i]=abias[i]/aweight[i]
abias.sort()
pyplot.scatter(x, abias)
\end{verbatim}

\subsection{TensorFlow and Heaviside}
If we look at the list of standadard activation functions (analogues of $\s$) offered by TensorFlow, we won't see Heaviside step function among them. Although it may be explained by difficulties for numeric differentiation, there still remains a question of the impact of smoothering into results of training.

To study that we take a network with 2D input and 2 layers having 3 and 30 nodes correspondingly, and custom sigmoid activation function
$$\frac{1}{1+\exp(-20x)}$$
being pretty close (in the uniform metric) to Heaviside.

We train it on 40 examples of the map:
$$[0,2]\times[0,2] \to \mathbb{R},\ (b,c)\mapsto b^2+c$$

\begin{figure}
\begin{center}
\includegraphics[scale=0.5]{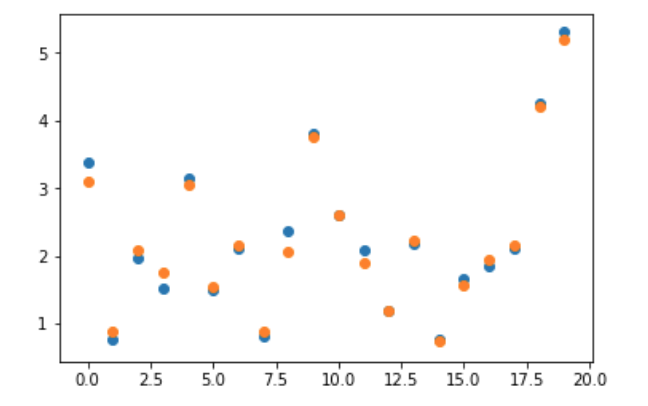}
\caption{({\it sample number, $f$-value}) plot for 20 samples of testing (blue) vs. prediction (orange) values produced by a 2-layer network trained on the map: $f(b,c)=b^2+c$}
\end{center}
\label{fig:quad_map_match}
\end{figure}

After 4000 iterations of passing through the set of training examples ($\text{epochs}=4000$) it produces pretty good matching (orange) with exact (blue) values on Figure 3.

Now we compute the output values produced by the network having the same values of $w$ and $b$ (which we obtained after training) but instead of sigmoid having Heaviside as activiation function. The results (orange) are compared to the output of our original network (blue) on Figure 4. We can see that even a slight smoothering results in significant discrepancy with Heaviside case already at the 2nd layer.

As a summary of experience with these examples,
we can say that using Heaviside step function enables us to approach
to analytic calculations of the expected structure (values of $W$ and $b$) of our network,
in particular - to  estimate the required number of layers and nodes in there,
deducing it from the algebraic structure of the transform/mapping/function being approximated by the network.
Numeric gradient descent calculations using smoothed versions of Heaviside result in blurring of the structure,
while in reward we are automatically (as a mere consequence of smoothness) getting an extrapolation of training values to the whole regions in between (which may look pretty impressive in cases when the approximated map really admits extrapolation).

\begin{figure}
\begin{center}
\includegraphics[scale=0.5]{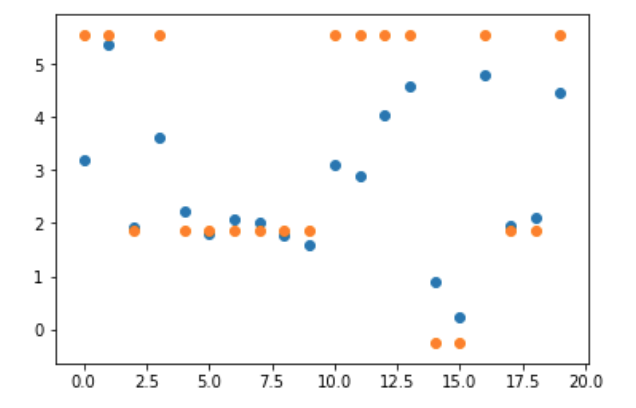}
\caption{({\it sample number, $f$-value}) plot for 20 samples of network output, trained  for the map $f(b,c)=b^2+c$ with sigmoid activation (blue), compared to the output computed for the same weights and biases, but with Heaviside activation (orange).}
\end{center}
\label{fig:quad_map_match}
\end{figure}

\section{Conclusion
\label{conc}}

To conclude, in this paper we made a next step in Heavisidisation program,
posed in \cite{2204.11613}.
We demonstrated that it is much more natural and easy-going than it could originally seem --
we explained how to get iterated-Heaviside analogues of elementary operations
and how to combine into answers for more serious problems.
Also we classified the ambiguities of Heavisidisation itself and additional
gauge freedom in its lifting to ML problem.

Now we can revisit the original questions about the epistemology ,
raised in  \cite{2204.11613},
and formulate them (just) a little better.
In short there are three groups of questions.

The first is quite constructive:

A) Heavisidisation of various problems and answers to them,
i.e. finding the analogues of the sides of (\ref{root2eq})
for more interesting constructions and theorems.

The other two are still conceptual:

B) What it means to find an {\it algebraic} formula
(through coefficients of $F(x)$)
for above answer, i.e. the r.h.s. of (\ref{root2eq}) for a given l.h.s.?
Whether and what one needs to assume about the shape of this answer?

C) How to distinguish between reasonable (acceptably short) and unreasonable answers?
Is there a {\it true} difference between these two notions?

The main question after all that is {\it what is the reasonable answer}?
One can hope to find, say, an answer for discriminant just as a polynomial in coefficients --
but it is infinitely complicated, not the one which we really wish.
Humans want/need something like (51) of \cite{0911.5278}, not the 2000-item expression
in the appendix to \cite{NLA} (book version).

But how can we distinguish between them without applying the human mind and taste?

And will a {\it computer} be ever able to distinguish?

Can the machine ever develop a {\it scientific taste}, similar to the human's one?

Does this human taste have any "objective" meaning or is just a pure reflection
of the phyical/biological structure of our brain?

Does this structure prevent us from discovering really deep laws of nature?

Or is it the one, exactly suited to complexity of these laws? \ \
(In religious terms, if the human's mind is exactly of the same kind  as Creator's)?

This set of questions is a standard dilemma in computer physics, see, for example,
a popular review in \cite{1308.4678}.

\bigskip

This paper teaches us that the answers to seemingly complicated questions
can actually be nearly trivial.
Perhaps, this is true also for the level (quality) of scientific answers.
Say, "deeper" answers can be associated with more layers
(iterations of Heaviside functions) --
at least this is an obvious difference between the two sides of
our sample example (\ref{root2eq}).
In fact, for a given number of layers the attraction orbits modulo gauge invariance
are relatively few.
They are invariants of the "group" which reshuffles nodes in a given architecture.
However what expresses scientific knowledge in this paradigm,
are {\it connections} between formulas at different levels,
and they can be more sophisticated and interesting.

\section*{Acknowledgements}

This work is partly supported by the Russian Science Foundation (Grant No.21-12-00400).

\end{document}